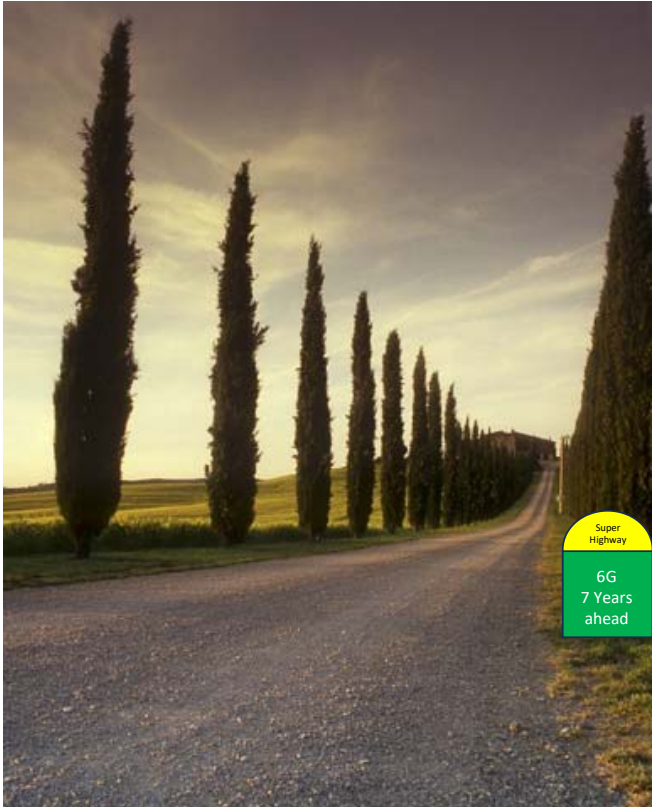

Super Highway

6G 7 Years ahead

# "MEETING IMT2030 PERFORMANCE TARGETS: THE POTENTIAL OF OTFDM WAVEFORM AND STRUCTURAL MIMO TECHNOLOGIES"

A white paper by IIT Hyderabad & WiSig Networks

## ABSTRACT

The white paper focuses on several candidate technologies that could play a crucial role in the development of 6G systems. Two of the key technologies explored in detail are Orthogonal Time Frequency Division Multiplexing (OTFDM) waveform and Structural MIMO (S-MIMO)

Table of Content

# Executive Summary: Meeting IMT2030 Performance Targets: The Potential of OTFDM Waveform and Structural MIMO Technologies

This executive summary offers a comprehensive look at the key 6G Candidate Technologies, with a primary focus on two essential technologies. Firstly, it delves into the objectives and framework of IMT-2030 (International Mobile Telecommunications for 2030 and beyond) established by the ITU (International Telecommunication Union). The discussion highlights various crucial goals, including inclusivity, ubiquitous connectivity, sustainability, innovation, enhanced security and privacy, resilience, standardization, interoperability, and interworking.

Two of the key technologies explored in detail are Orthogonal Time Frequency Division Multiplexing (OTFDM) waveform and Structural MIMO (S-MIMO).

OTFDM is proposed as a candidate for 6G mobile communication, addressing the limitations of traditional OFDM waveform. OTFDM enables simultaneous transmission of reference signals (RS) and data with low peak-to-average ratio (PAPR), high power efficiency, and reduced overhead. The benefits, technical details, simulation results, and potential applications of OTFDM are discussed.

S-MIMO revolutionizes network capacity by implementing highly directional beams and structural arrangements of multiple antenna panels, enabling 360-degree coverage in both azimuth and elevation. This technology overcomes limitations of conventional 5G MIMO systems and offers significant capacity improvements. The experimental results and different structural arrangements are presented. ***The combination of OTFDM and S-MIMO technologies are shown to have the potential to meet the peak data rate, hyper latency, spectrum efficiency and power efficiency requirements set forth by IMT-2030 Framework document.***

The white paper also highlights the importance of enhancing coverage with relays and explores the interworking between terrestrial and satellite communications (Satcom) for seamless connectivity, particularly in challenging geographical areas. Furthermore, the integration of Artificial Intelligence (AI) and Machine Learning (ML) into the IMT-2030 ecosystem is discussed, focusing on advancements in the physical layer, MAC and higher layers, and the potential for advanced data analytics.

Lastly, the white paper introduces the concept of integrated sensing and communication, which enables precise positioning with centimeter-level accuracy through wireless methods. The benefits of this integration in industrial environments are emphasized.

Overall, this white paper provides valuable insights into the technologies that can shape the future of 6G mobile communication systems. It offers a comprehensive analysis of the candidate technologies, their applications, and their implications for future wireless communication networks.

## Objectives and Framework of IMT-2030

The ITU WP 5D recommendation describes the framework and overall objectives for the development of the terrestrial component of International Mobile Telecommunications (IMT) for 2030 and beyond (IMT-2030). IMT-2030 (a.k.a. 6G) aims to better serve the needs of the networked society in both developed and developing countries. The framework includes a broad variety of capabilities associated with envisaged usage scenarios and addresses the objectives for the development of IMT-2030, including enhancement and evolution of existing IMT and aspects of interworking with other networks.

The Recommendation highlighted several goals that IMT-2030 aims to achieve, including:

| Goals | Aims to Achieve |
|---|---|
| Inclusivity | Bridging the digital divide and ensuring meaningful connectivity for everyone |
| Ubiquitous connectivity | Providing affordable connectivity and basic broadband services with extended coverage, including sparsely populated areas. |
| Sustainability | Building on energy-efficient, low power consumption technologies to address climate change and contribute to sustainable development goals |
| Innovation | Fostering technological advances that improve user experience, productivity, and resource management. |
| Enhanced security, privacy, and resilience | Designing the future IMT system to be secure, privacy-preserving, and capable of quick recovery from disruptive events. |
| Standardization and interoperability | Designing systems with standardized and interoperable interfaces to ensure compatibility and cooperation between different network parts. |
| Interworking | Supporting service continuity and flexibility through close interworking with non-terrestrial network implementations, existing IMT systems, and other access systems. |

## Applications and Services Enabled by IMT-2030

The following Table summarizes the applications and services enabled by IMT-2030.

| Application | Description |
|---|---|
| Ubiquitous intelligence | AI and ML spread throughout communication, enabling smart cities and autonomous network management |
| Ubiquitous computing | Data processing expands to cloud & devices, support real-time responses, E2E AI applications |
| Immersive multimedia & multi-sensory interactions | Personalized Extended Reality (XR) and human-machine interfaces for remote operations |
| Digital twin and virtual world | Create digital twins, impacting industries like healthcare, agriculture, and construction |
| Smart industrial applications | Scale smart industrial applications with real-time intelligence and intelligent device connectivity |
| Digital health and well-being | Enhances digital health services and enables pervasive IoT devices |
| Ubiquitous connectivity | Connect underserved areas, bridging the digital divide with consistent user experience |
| Integration of sensing & communication | Integrate sensing for innovative applications and enhanced situational awareness |
| Sustainability | Aim for environmental, social, economic sustainability, leveraging circular economy principles |

Table-1:IMT-2030 Applications

## Usage Scenarios and Capabilities of IMT-2030

Figures 1 and 2 describe the six usage scenarios and their associated capabilities. Notably, India played a crucial role in shaping the IMT-2030 framework, introducing the "Ubiquitous Communications" scenario as the sixth one. The Indian delegation, led by DOT/TEC/WPC, collaborated with academia and industry partners such as IITH, Tejas Networks, IITM, CEWiT, Reliance Jio, Sankhya Labs, WiSig Networks and IISC, making significant contributions during the meetings. India's introduction of the "Ubiquitous connectivity" scenario, along with enhanced coverage capabilities, paved the way for 6G technology development, offering affordable connectivity and broadband services to rural and sparsely populated areas. This strategic move will help bridge the digital divide and create economic opportunities for India's vast rural population.

### Figure 1 - 6 Usage Scenarios of IMT-2030

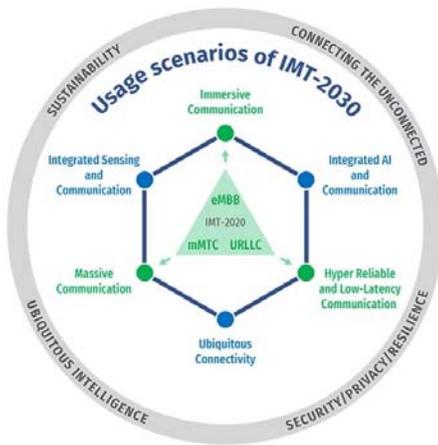

**This is called Wheel Diagram**

**Extension from IMT-2020**
- eMBB ➔ **Immersive Communication**
- mMTC ➔ **Massive Communication**
- URLLC ➔ **HRLLC**

**Newly Added**
- **Ubiquitous Connectivity**
- **Integrated Sensing and Communication**
- **Integrated AI and Communication**

**Four Overarching Aspects**
- **Connecting the Unconnected**
- **Security / Privacy / Resilience**
- **Ubiquitous Intelligence**
- **Sustainability**

### Figure 2 – Capabilities of IMT-2030

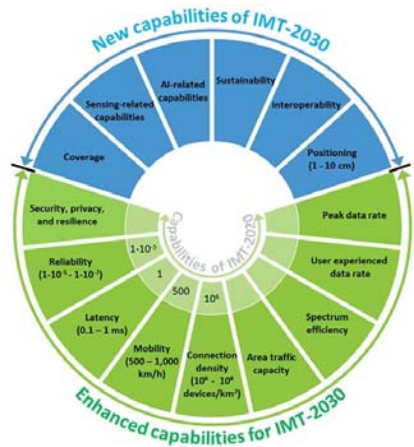

**This is called Palette Diagram**

**The range of values given for capabilities are estimated targets for research and investigation of IMT 2030**

**All values in the range have equal priority in the research and investigation**

**For each usage scenario, a single or multiple values within the range would be developed in future in other ITU-R recommendations / reports**

Furthermore, India presented another critical capability named "Interface interoperability," aiming to promote open interfaces among system entities and facilitate the deployment of interoperable radio access network systems. This approach will not only expedite the delivery of next-gen 6G and beyond wireless infrastructure to operators but also foster innovation and open market competition among Indian suppliers/product companies. The adoption of open interfaces allows smaller vendors and operators to introduce customized services, tailoring the network to suit their specific requirements.

Additionally, India's emphasis on "Energy efficiency" brought a sustainability aspect to the IMT-2030 document. This focus highlights the importance of low-power consumption technologies, reduced greenhouse gas emissions, and resource utilization under the circular economy model, contributing to climate change mitigation and sustainable development goals.

Following the publication of the IMT-2030 Framework in the 5D WP #44, the projected work plan for WP5 over the next 12 months includes the discussion on Test Performance Requirements (TPR) and the mapping of Usage Scenarios with Capabilities in 5D WP #45 (Feb 2024). Subsequently, in 5D WP #47 (Oct 2024), the discussions on Evaluation Methods and Test Environment will commence, with the actual TPR values for different test environments being developed during the 2025 timeframe (see Figure-3 for the tentative timeline of the entire IMT-2030 process).

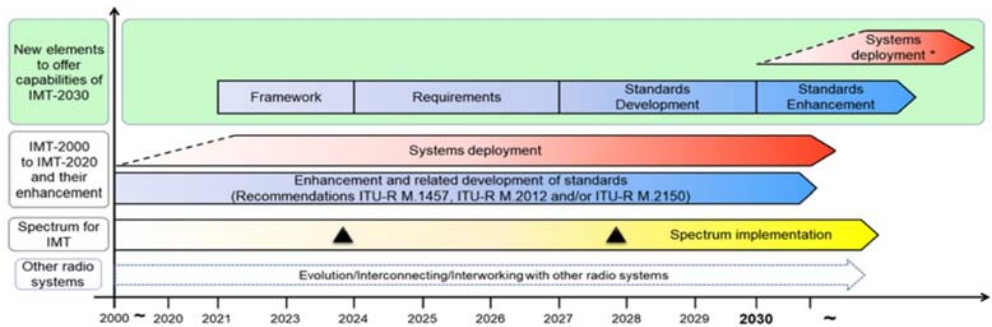

Figure 3 – Relationship & Timelines of IMT-2030

To significantly contribute to 6G standards development, India is encouraged to have technical concepts, system architecture, initial prototypes, and early simulation results available as soon as possible. The subsequent sections will introduce new technologies and their associated TPRs, which hold the potential to become integral components of the 6G standards.

# IMT-2030 Candidate Technologies

## OTFDM Waveform

Orthogonal Frequency Division Multiplexing (OFDM) has been a fundamental technology in modern wireless communication systems, such as WLAN, 4G, and 5G, due to its spectral efficiency, multi-user capability, and ease of channel estimation. However, the traditional OFDM waveform has inherent limitations, including a high peak-to-average ratio (PAPR) and power inefficiency.

We introduce Orthogonal Time Frequency Division Multiplexing (OTFDM) as a candidate for 6G mobile communication. OTFDM is designed to address the shortcomings of traditional OFDM and offer a waveform that enables simultaneous transmission of reference signals (RS) and data with low PAPR, high power efficiency, and reduced overhead.

The conventional downlink of cellular systems often experiences high PAPR due to multiple users sharing spectral resources. Consequently, power amplifiers (PA) require significant back-off, resulting in poor power efficiency and increased complexity. While 5G New Radio (NR) introduced DFT-S-OFDM with lower PAPR for the uplink, the adoption of OFDM as the unified standard waveform for both sub-6 GHz and mmWave services limited the power efficiency.

To achieve the ambitious goals set by the International Telecommunication Union (ITU) for 6G, including extremely low latency, high data rates, and power efficiency, a new waveform is needed. OTFDM presents an innovative solution by time-multiplexing RS and data within a single OFDM symbol, thereby allowing one-shot transmission with flexible RS overhead and high-power efficiency.

The OTFDM symbol is generated by the following set of operations: RS and Data/Control Information is Multiplexed into a single sequence and is fed to a DFT module followed by excess bandwidth spectrum shaping at subcarrier level further followed by IFFT and CP addition operation which generates a OTFDM symbol". By expanding the bandwidth and shaping the spectrum using a pulse shaping filter, the adverse effects of inter-symbol-interference (ISI) are minimized, in addition to reducing the PAPR. The design parameters of RS density, excess bandwidth, and DFT size can be carefully selected to eliminate the irreducible error floor caused by the channel's impulse response.

Through extensive computer simulations based on 3GPP channel models, we demonstrate the effectiveness of OTFDM in performing self-contained channel estimation in typical delay spread channels for modulation sizes up to 256 QAM. The provided PAPR results for different modulation schemes indicate that OTFDM achieves low PAPR operation and high PA efficiency, making it a promising candidate for 6G mobile communication.

In the following sections of this paper, we delve into the technical details of OTFDM, present simulation results, and discuss its applications and implications for future wireless communication systems.

OTFDM Symbol Generation

This includes the following operations as shown in the Figure:

1. **Time multiplexing**: The central idea is to time multiplex data/control information and the associated reference signal (RS) in a block called OTFDM symbol (See Figure 4). This operation enables self-contained channel estimation, data/control decoding with least possible latency. The RS is appended by RS CP (RS circular pre-fix) and RS CS (RS circular suffix) that will act a guard between RS and data and also enables FFT based channel estimation.

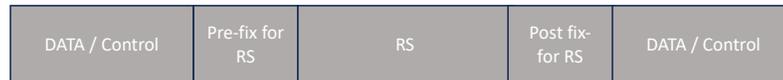

Figure 4: OTFDM Symbol Structure

| DATA / Control | Pre-fix for RS | RS | Post fix-for RS | DATA / Control |

2. **DFT precoding**: As shown in Figure-5, the multiplexed time sequence of length M is fed to a M-point DFT module

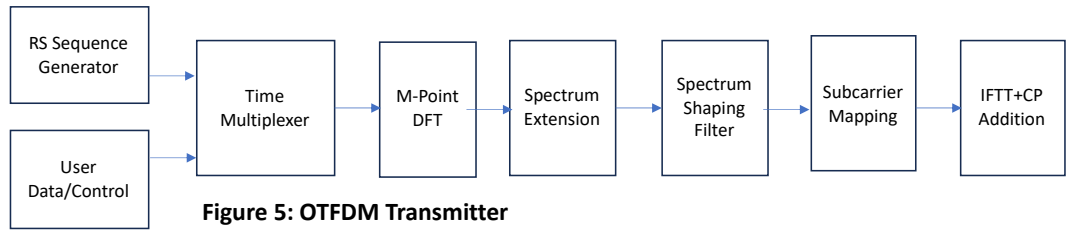

Figure 5: OTFDM Transmitter

3. **Bandwidth Expansion and Spectrum shaping**: The output of the DFT is appended with a prefix (last 'v' samples of the DFT output) and suffix (first 'v' samples of the DFT output). A frequency domain spectrum shaping filter is applied to spectrum extended sequence before applying a subcarrier mapping operation. The excess bandwidth is "2v" subcarriers. The frequency domain shaping filter is chosen to meet Nyquist Criterion for Zero ISI i.e., the folded squared spectrum of the spectrum shaping filter is constant over the M-subcarriers of interest (See Figure-6).

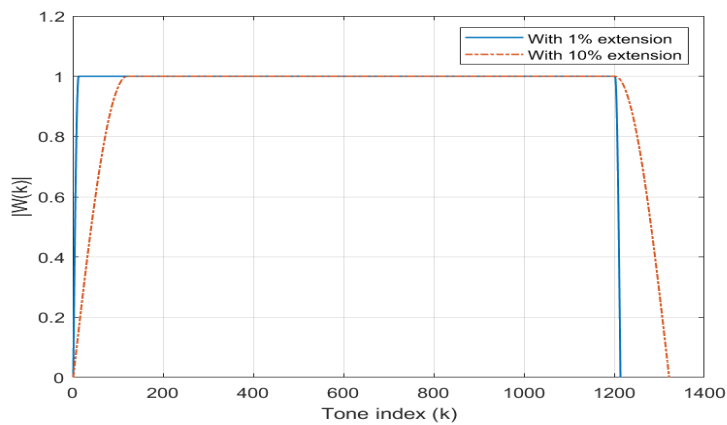

Figure 6: An example frequency domain spectrum shaping filter

4. **Time Domain Signal Generation**: IFFT followed by CP addition is performed on the subcarrier mapped sequence to generate an OTFDM signal.

The receiver performs channel estimation and equalization on per OTFDM symbol basis.

**Hyper Low Latency**

The physical layer latency is determined by the transmission time interval (TTI). OTFDM has the ability to transmit and receive data/control information using one OTFDM symbol which is the TTI duration. Table-2 shows the TTI duration required for different subcarrier spacings (SCS). This parameter gives an indication of SCS required to design a hyper low latency system.

| Subcarrier Spacing | 30 KHz | 60 KHz | 120KHz | 240KHz | Units |
|---|---|---|---|---|---|
| System bandwidth | 100 | 200 | 400 | 800 | MHz |
| FFT size | 4096 | 4096 | 4096 | 4096 | |
| Sampling time | 8.138 | 4.069 | 2.0345 | 1.01725 | n Seconds |
| Symbol duration | 33.33 | 16.67 | 8.33 | 4.167 | μ Seconds |
| CP duration | 2.344 | 1.172 | 0.586 | 0.293 | μ Seconds |
| TTI (symbol + CP duration) | 35.677 | 17.839 | 8.919 | 4.459 | μ Seconds |

Table-2: OTFDM TTI duration for different subcarrier spacings (SCS)

**Peak-To-Average-Power Ratio (PAPR)**

The PAPR of the waveform is related to PA back-off and power efficiency of the waveform. The CCDF of the PAPR is shown in the Figure 7-9 as a function of the excess BW. DFT-S-OFDM is specified in 5G NR standard as a mandatory feature for UE and optional mode for the gNB. The PAPR of this waveform is shown in Figure 7. In this case, pi/2 BPSK applies a two equal tap spectrum shaping filter without bandwidth expansion.  3GPP RAN4 evaluations show that, for the RAN4 specified ACLR and EVM targets, the pi/2 BPSK with spectrum shaping operates close to the PA saturation power whereas QPSK requires 3.0 dB PA back-off. The PAPR results of Figures 7-9 show the further PAPR reduction achieved by the OTFDM waveform as a function of excess BW used in signal generation. The reduction in PAPR can be further exploited by the system to reduce the PA back-off those results in higher transmission power, higher coverage and higher energy efficiency.

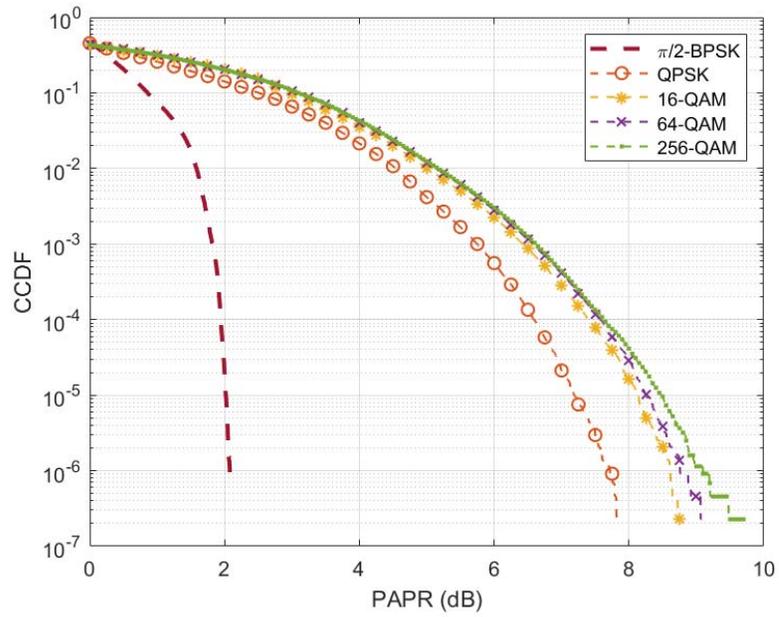

Figure 7: PAPR of Rel-15 5G-NR DFT-S-OFDM signals with different modulations

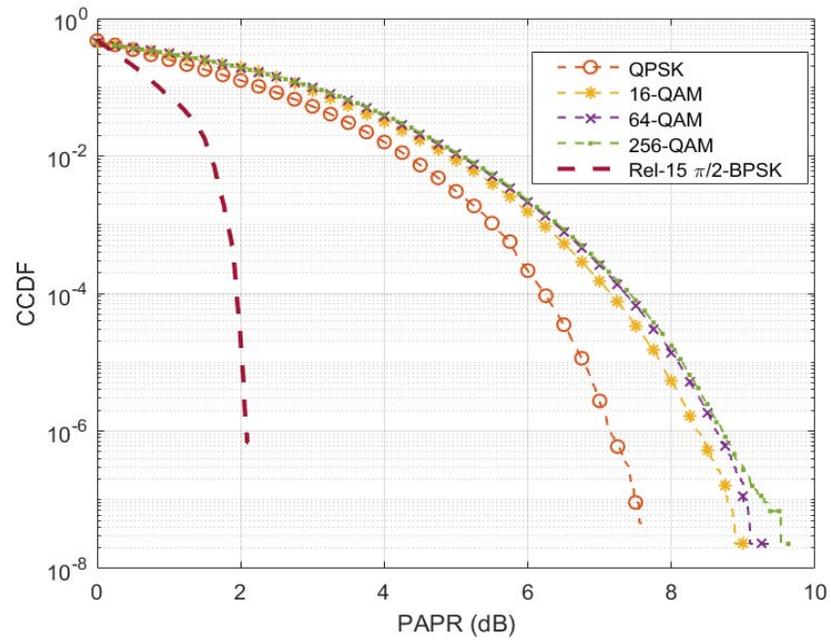

Figure 8: PAPR of OTFDM with different modulations and 5% spectrum extension

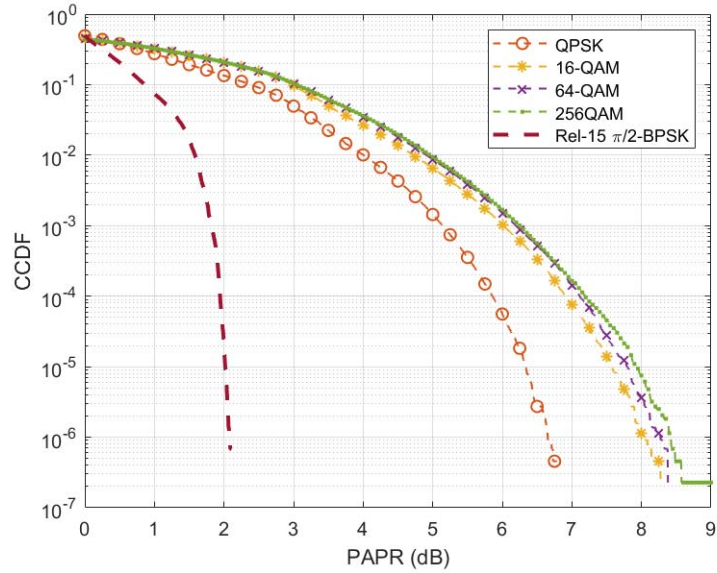

Figure 9: PAPR of OTFDM with different modulations and 10% spectrum extension

**RS Overhead**

The receiver performance of OTFDM is determined by the RS overhead, RS CP size, and the excess BW of the spectrum shaping filter. Table 3 shows the RS+RS CP+RS CS requirement for 120KHz SCS to ensure that the receiver does not experience error floors and the BLER performance is acceptable. Table 4 shows the simulation parameters used in conducting link level evaluation and Figure 10 shows the BLER performance for different modulation types. It is observed that with a properly designed RS and the associated CP, channel estimation and equalization can be performed on per-symbol basis.

| 198 PRBs with 5% extension | | | | | |
|---|---|---|---|---|---|
| Modulation | R | RS (samples) | RS-CP or RS CS (samples) | Total overhead (RS+RS CP+RS CS) (samples) | % of overhead |
| QPSK | R = 0.44 | 197 | 47 | 291 | 12 |
| 16 QAM | R = 0.49 | 197 | 47 | 291 | 12 |
| 64 QAM | R = 0.6 | 197 | 97 | 391 | 16.5 |
| 256 QAM | R = 0.8 | 197 | 97 | 391 | 16.5 |

Table-3: RS overhead requirement for each modulation scheme with 120KHz subcarrier spacing

**Simulation Parameters**

| Parameter | Value |
|---|---|
| SCS | 120KHz |
| No. of Tx antennas | 1 |
| No. of Rx antennas | 1 |
| Number of PRBs | 200 (M = 2400) |
| Excess BW | 5% (10 PRBs) |
| Number of symbols | 1 |
| Modulation | QPSK, 16 QAM, 64 QAM, 256 QAM |
| RS size | 139 |
| RS CP + RS CS size | 67*2 = 134 |
| Coding rate | 0.4385, 0.478, 0.852, 0.89 |
| Channel model | TDL-D 10 nsec |

Table 4: Simulation parameters

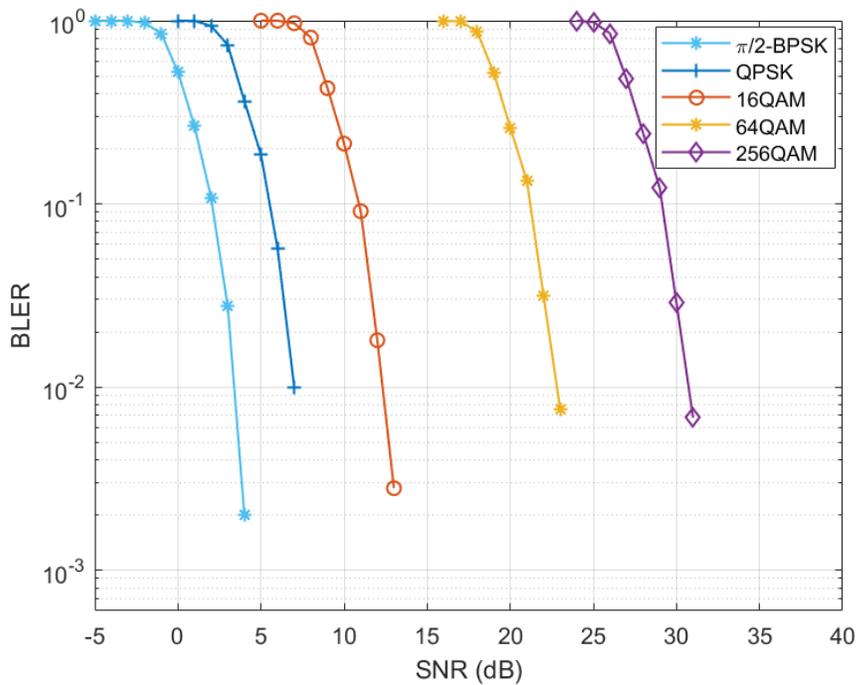

Figure 10: BLER performance with OTFDM waveform in TDL-D

In summary, OTFDM is shown to be an excellent candidate in meeting the Hyper Latency, High Power Efficiency and High Data rate requirements of IMT 2030 applications. Further evaluation of this waveforms will be conducted with prototype systems operating in the target frequency bands of IMT 2030 systems.

## Revolutionizing Network Capacity with Structural MIMO (S-MIMO)

Traditional 5G networks rely on two-dimensional (2D) antenna arrays, featuring antenna elements distributed in both azimuth (horizontal) and elevation (vertical) dimensions. These arrays consist of patch antennas or sub-arrays, referred to as ports, each providing sectoral coverage of 120 degrees in azimuth and approximately 60 degrees in elevation. In 5G MIMO deployments, up to 64-antenna ports are utilized. These deployments typically involve three sectors, each with a 64-port 2D rectangular array operating at a carrier frequency of around 3.5 GHz. To achieve a significant increase in network capacity, the base station employs precoding or beamforming in the downlink (DL), serving multiple users on the same time-frequency slots. The selection of users for multi-user MIMO (MU-MIMO) transmission is based on spatial separation and channel state information (CSI) of active users, which can be obtained through various methods such as CSI feedback, estimations on sounding reference signals (SRS) in the uplink (UL), and TDD reciprocity calibration.

In the DL, the base station pairs users based on DL CSI and formulates precoder weights for the paired users, while in the UL, channel estimations on the reference signals associated with the 64-antenna ports are used to combine the received data in the uplink equalizer. However, due to the limited directivity resulting from the low number of antenna elements per port, the signal power received at each port in the uplink is low. This leads to poor channel estimation quality, particularly for users at cell boundaries (cell-edge users) or for SRS channel estimation. Consequently, DL precoder weights are prone to errors, especially for cell-edge users. This bottleneck hampers the maximization of network throughput with MU-MIMO.

In each sector, using 64 antenna ports and DL MU-MIMO, the base station can serve up to 16 layers, in a paired MU-MIMO group. Although a theoretical 16-fold increase in network capacity can be expected compared to a single-antenna system, practical limitations such as limited antenna directivity per port, interference from other sectors, and power division among scheduled layers restrict the number of users that can be paired and the achievable network capacity.

To overcome these limitations, we propose Structural MIMO (S-MIMO), which eliminates the constraints of conventional 5G MIMO systems by implementing the following methods:
- Highly directional beams associated with each antenna port/panel maximizes the coverage per Port/Panel
- Structural arrangements of multiple antenna panels using a three-dimensional structure, enabling full coverage in azimuth and elevation.
- Joint baseband processing of signals associated with multiple antenna panels, incorporating TDD reciprocity calibration, MU-MIMO precoding/beamforming in the DL, and joint processing of UL signals from multiple antenna panels.

S-MIMO involves multiple antenna panels at the base station, each equipped with multiple antenna elements and offering horizontal and vertical antenna spacing. This configuration allows for the physical orientation of signals in both azimuth and elevation directions. Each antenna panel can feature multiple antenna ports, with each port associated with a highly directional static beam in azimuth and elevation. By structuring a group of such antenna panels, the multiple beams from the multiple antenna ports collectively cover a full 360 degrees in azimuth and 180 degrees in elevation. With S-MIMO, the base station can provide cellular coverage to users on the ground, as well as drones and unmanned aerial vehicles in the sky, ensuring ubiquitous coverage. The employment of multiple highly directional beams in azimuth and elevation enables significant capacity improvements.

The realization of the S-MIMO structure can take various forms, such as implementing it atop regular cellular towers, on the facade of buildings, or utilizing large balloon-like structures in the sky, featuring configurable spacing and orientation of antenna panels to full coverage. All these antenna panels are connected to a centralized baseband processing unit, where signal processing for transmitted and received signals occurs. This baseband processing includes multi-user MIMO precoding/beamforming, joint processing of UL signals, TDD reciprocity calibration between signals of multiple antenna panels, and more. Different structural arrangements of antenna panels are explored to enhance coverage and capacity.

**Experimental Results**

Structural Arrangements

Single 30-degree Panel: Figure 11 shows a single antenna panel comprising 192 antenna elements and 48 antenna ports, where 4 antenna elements in a row are combined to form a port/beam with a 30-degree coverage in azimuth and 90-degree coverage in elevation, with 12 dB directivity and 30 dBm transmit power per port. The panel consists of 12 ports placed horizontally and 4 ports placed vertically. The antenna panel is located at a height of 2.5 meters as shown in the Figure 11. The UEs are placed at a distance of 30 meters as shown in the Figure 12 where UEs are distributed throughout the building in azimuth and elevation. Each UE has two antennas and there 18 UEs with a total of 36 antennas. The UEs are placed to cover 30-degree span in azimuth with respect to the transmitter. The antenna ports of the transmitter are connected a number of RRHs that are connected to a pool of DSPs by fiber. The experiment uses 15KHz numerology, the physical layer of the transmitter and receiver are designed to support MU MIMO operation in the DL. User pairing and MU MIMO precoding is performed every TTI in the centralized baseband DSP processor. The experiment showed that out of 36 UE antennas, 10 layers are scheduled consistently every TTI where each layer has QPSK modulation.

This experiment shows that high directivity narrow beam with 30-degree azimuth coverage can pair an average of 10 layers with up to 2 bits/sec/Hz per layer. This remarkable capacity increase becomes even more pronounced when multiple such panels are employed to achieve 360-degree coverage, leading to an exceptional enhancement in overall capacity.

Figure 11: S-MIMO Transmitter

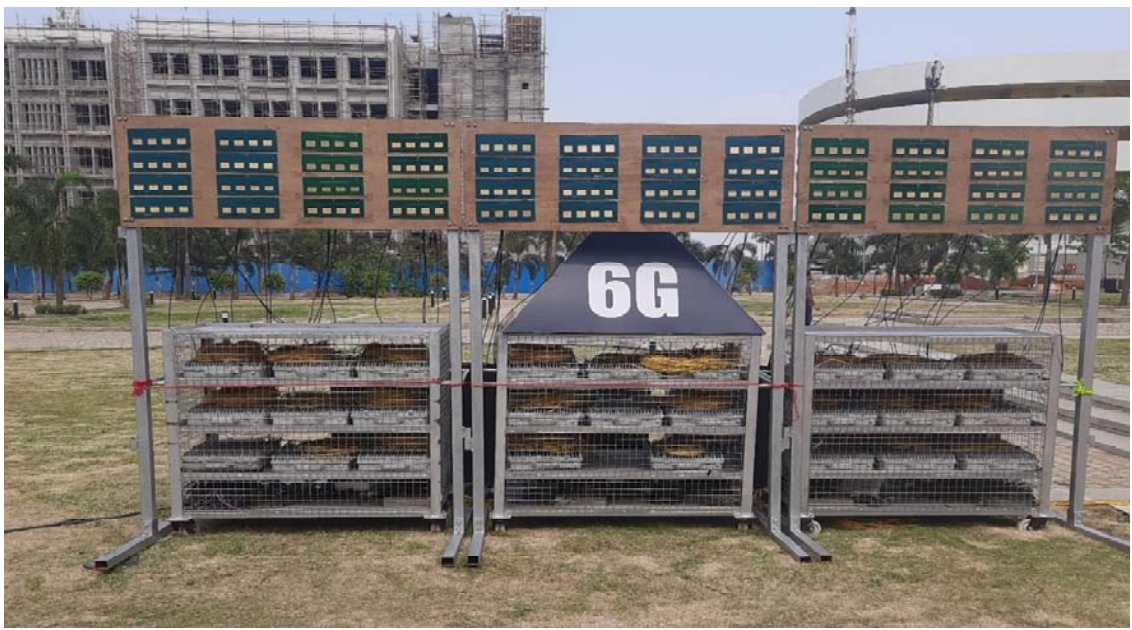

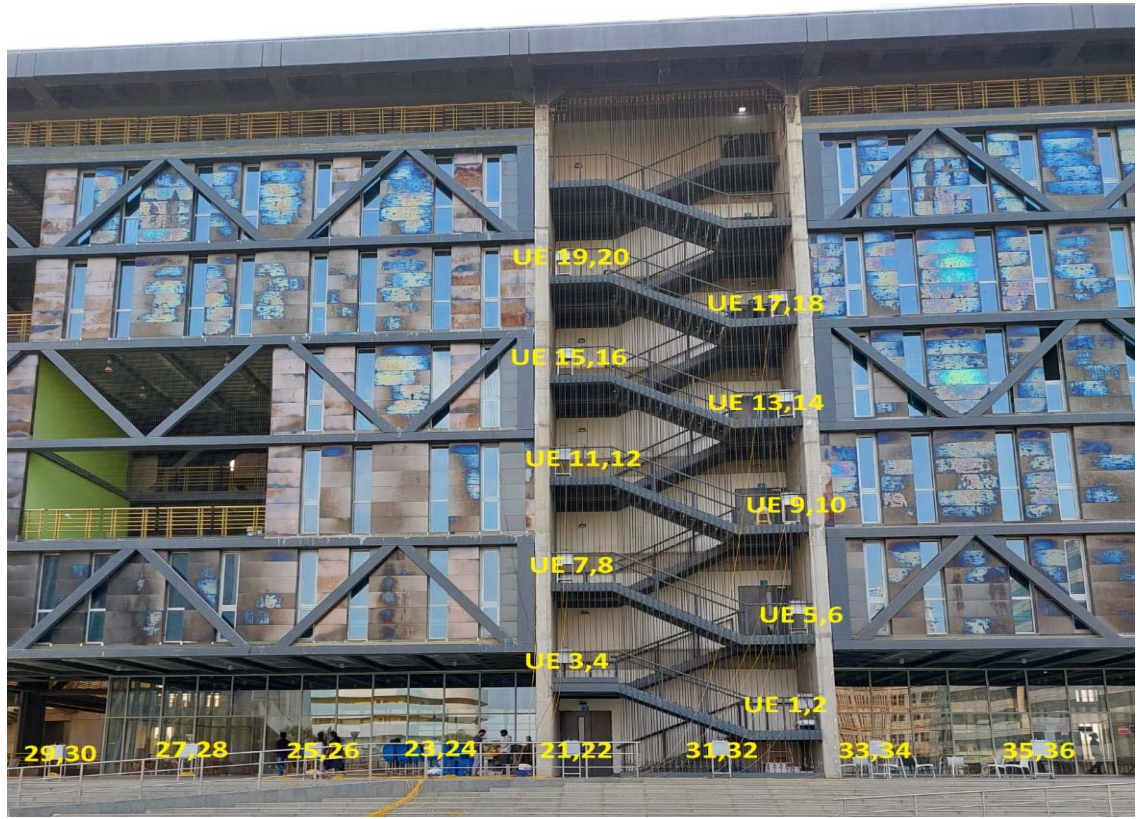

Figure 12: UE Placements in a 30-deg beam region

**Network Level Evaluation of S-MIMO concept**

Building upon the compelling outcomes of our prior experiments, we proceeded to conduct a network-level performance evaluation of the S-MIMO concept using simulation. In this evaluation, we employed 3, 6, and 12 antenna panels to achieve comprehensive 360-degree coverage. The spectrum efficiency results obtained from this evaluation are presented in Table 5 and Figure 13.

Remarkably, the results reveal that S-MIMO can achieve up to 116 Bits/Sec/Hz (all Panels combined), showcasing a significant capacity increase when compared to existing 5G systems that offer up to 21 bits/sec/Hz, representing a remarkable 5.5-fold spectrum efficiency improvement.

| Antenna Panels | No of Tx Ports /Panel | No of Antenna elements all Panels combined | 5% SE | Mean UE SE | Average SE/Panel (Bits/sec/Hz) | Total SE in 360-deg (Bits/sec/Hz) |
|---|---|---|---|---|---|---|
| 3 | 32 | 192 | 0.040 | 0.694 | 6.94 | 21 |
| 6 | | 1152 | 0.161 | 0.860 | 9.03 | 54 |
| 12 | | 4608 | 0.172 | 0.992 | 9.66 | 116 |

Table 5: DL S-MIMO MU MIMO SE with multiple antenna panels – Mid-band

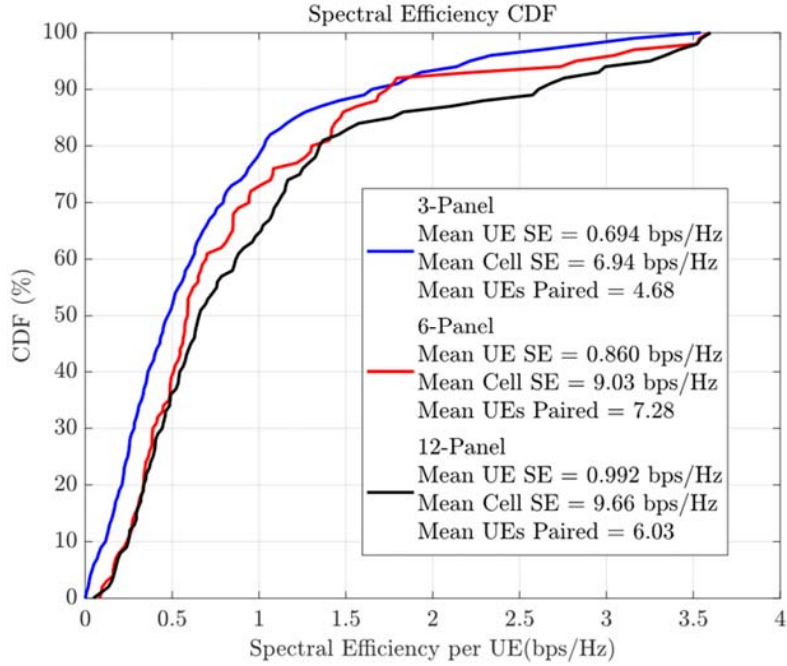

Figure 13: DL S-MIMO MU MIMO SE CDF per Panel Results in Mid-band

DL S-MIMO MU-MIMO simulation assumptions

| Parameters | | Description |
|---|---|---|
| Carrier frequency ($f_c$) | | 3.5 GHz |
| Total number of ports at BS | | 32 |
| Antenna array at BS | 3 antenna panels | $M \times N \times P = 4 \times 8 \times 2$<br>K=2 (Number of elements per port in vertical)<br>L=1 (Number of elements per port in horizontal)<br>Total number of antenna elements per port = 2<br>Number of polarizations P = 2<br>120-degree patten in azimuth<br>Boresight angles = [0 120 240] |
| | 6 antenna panels | $M \times N \times P = 4 \times 24 \times 2$; K=2; L=3<br>Total number of antenna elements per port = 6<br>60-degree pattern in azimuth<br>Boresight angles = [0 60 120 180 240 300] |
| | 12 antenna panels | $M \times N \times P = 4 \times 48 \times 2$; K=2; L=6<br>Total number of antenna elements per port = 12<br>30-degree pattern in azimuth<br>Boresight angles = [0 30 60 90 120 150 180 210 240 270 300 330] |
| Antenna array at UE | | **$M \times N \times P = 1 \times 2 \times 2$**<br>Total number of ports at UE = 4 |
| Basic element pattern | | 120-degree patch antenna with maximum directivity of 8dB |

| Channel Model | Urban Macro |
|---|---|
| BS layout | Hexagonal layout with 1-ring (7 hexagonal cells in a ring) |
| Inter-site distance (ISD) | 500m |
| Transmit Power ($P_t$) | 43dBm |
| Bandwidth | 10MHz |
| Subcarrier spacing | 15KHz |
| BS height | 25m |
| UE height | 1.5m (Indoor UE height = [1.5 22.5]m) |
| Average UEs attached per panel | 10 |
| Maximum UEs paired for MU-MIMO per panel | 8 (UE pairing enabled) |
| Hybrid ARQ | Maximum HARQ processes = 10<br>Outer loop rate control enabled (10% BLER) |
| Scheduler | Proportional fair scheduler |
| MU-MIMO Precoding | Regularized Zero Forcing based on UL SRS CSI |
| Control overhead | 31% |

## IMT 2030 Capabilities obtained using a fusion of OTFDM and S-MIMO

We undertake an assessment to ascertain the attainability of IMT-2030 key radio network capabilities by leveraging a fusion of OTFDM and S-MIMO principles. The ensuing passage presents a concise overview of our analysis.

- **Hyper Low Latency:** OTFDM allows transmission of data/control and RS in a single shot transmission. Latency is primarily dictated by the TTI duration, which is 17.839, 8.919 and 4.49 micros seconds for 60,120, and 240 KHz SCS, respectively. In practice, considering the scheduling latency and the latency due to the TDD frame structure, the overall processing time required to achieve for packet transmitted to reach the receiver can meet the IMT 2030 targets.

- **Waveform Efficiency:** The OTFDM waveform can reach close to 100% PA efficiency for pi/2 BPSK case to serve users at the cell edge. Higher order modulation alphabets can be used to serve users located closer to the base station.

- **Spectrum Efficiency:** IMT-2030 framework document envisages a 2-3 fold increase in average spectrum efficiency over 5G NR. The 12 Panel S-MIMO performance meets/exceeds this target.

- **Coverage:** The Link Budget Table illustrates the coverage in a Urban Macro scenario, where a 384-antenna element panel is utilized. The DL coverage is shown in Figures 14 (Urban Macro) and 15 (Rural Macro) for 7 GHz frequency band (with 250MHz allocated bandwidth), indicating the feasibility of achieving coverage comparable to midband using S-MIMO concepts. However, it is important to note that UL coverage faces limitations due to the constrained availability of antennas and power at the UE as shown in Figures 16 and 17 for Urban Macro and Rural Macro respectively.

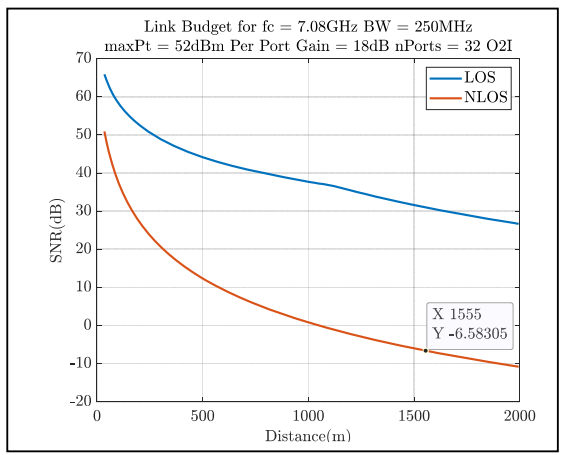

Figure-14: DL Link Budget for indoor UEs (UMa)

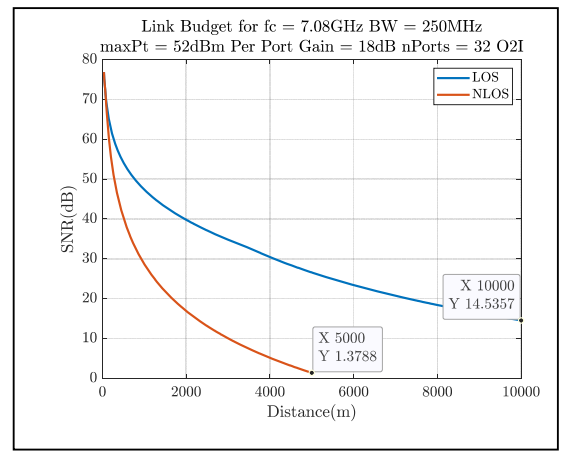

Figure-15: DL Link Budget for indoor UEs (RMa)

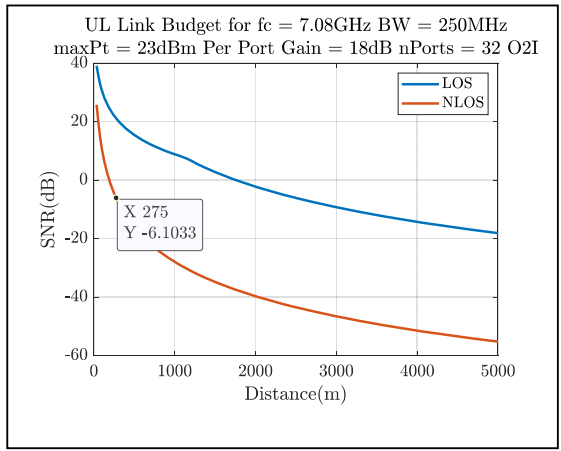 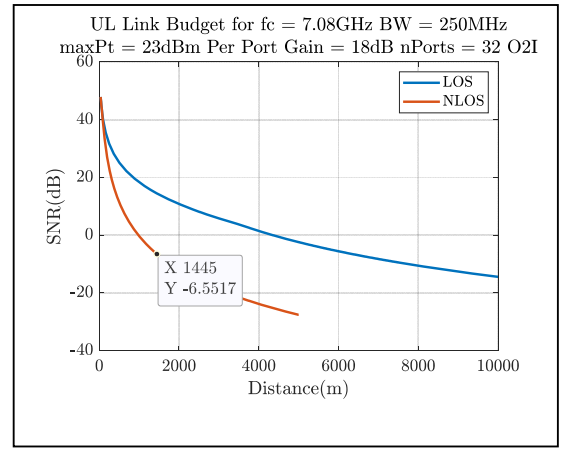

Figure-16: UL Link Budget for indoor UEs (UMa)     Figure-17: UL Link Budget for indoor UEs (RMa)

Link budget simulation assumptions

| Parameters | Description |
|---|---|
| Carrier frequency ($f_c$) | 7.075 GHz |
| Total number of ports at BS | 32 |
| Maximum BS port gain | 18dB |
| Channel Model | Urban Macro, Rural Macro |
| Transmit Power ($P_t$) | DL: 52dBm (5Watt per port) UL: 23dBm |
| Bandwidth | 250MHz |
| BS height | 25m |
| UE height | 1.5m |
| Noise power density | -174dBm/Hz |
| Indoor losses | As per TR 38.901 |

To address these constraints in upper mid-bands, subsequent sections elaborate on innovative network topologies such as relays and IAB, which offer promising solutions for UL coverage requirements.

- **Peak Data Rate:** The value of this parameter relies on the system's employed bandwidth. In the upper midband, where 1GHz is assumed to be available in a region with four operators, a minimum bandwidth of 250MHz can be allocated to each operator. This increase in bandwidth alone leads to a remarkable enhancement of at least 2.5 times in average, cell-edge, and peak data rates.

The comprehensive analysis and ensuing results conclusively demonstrate that the amalgamation of OTFDM and S-MIMO technologies exhibits the potential to fulfil the targeted requirements set by ITU WP 5D.

### Further Enhancements

The following sections outline further enhancements that are feasible for IMT 2030 systems.

### Enhancing Coverage with Relays

In traditional cellular configurations, the gNB is situated on a dedicated tower, transmitting signals with sufficient power to achieve the desired coverage. However, the UE faces limitations on its transmission power due to SAR limits, and it is challenging to control the directionality and orientation of antennas. As IMT 2030 explores higher frequency bands, achieving high data rates in the uplink becomes increasingly difficult. Research on IMT 2030 may explore network topologies that utilize relays to enhance both network coverage and capacity. In 3GPP 5G NR Rel-17, the Integrated Access Back-haul (IAB) concept has been defined, allowing a combination of mmWave and sub-6 GHz radio to form a chain of relays. The IAB concept can be further improved by:

1. Adopting a power-efficient OTFDM waveform and utilizing directional beams through S-MIMO concepts to increase coverage.
2. Mitigating high propagation losses associated with mmWave by using new frequency bands in the upper mid band range as an alternative.
3. Introducing innovative network topologies that involve relays mounted on streetside poles, MIMO structures placed on building facades, high altitude platforms serving as relays, and IAB nodes.
4. Exploring Next Gen IAB, which enables hyper low latency and highly reliable communication, can minimize end-to-end latency, maximize reliability, and further enhance data rates. For rural and hard-to-reach areas, a chain of relays with a range of hundreds of kilometers should be considered.

### Interworking Between of Terrestrial and Satcom

India encompasses vast regions of rugged terrain and dense forests, where conventional connectivity options are limited. To overcome this challenge, a combination of terrestrial and satellite networks is indispensable. Satellite communications (Satcom) play a crucial role in facilitating connectivity during natural calamities and disasters. Additionally, the utilization of low bit rate Satcom enables a multitude of Internet of Things (IoT) applications in rural and underserved areas. To advance this field, the following Satcom technologies warrant further investigation:

1. Direct-to-Satellite Narrowband Internet of Things (NB IoT) for Geostationary Earth Orbit (GEO).
2. High bit rate connectivity leveraging Low Earth Orbit (LEO) constellations.

The 3rd Generation Partnership Project (3GPP) 5G New Radio (NR) release 17 has already addressed the aforementioned use cases. Nevertheless, there is a need for additional enhancement of NB IoT functionality, particularly in operating at lower Signal-to-Noise Ratios (SNRs). This enhancement would enable direct connectivity to GEO Satellites. Furthermore, the interworking between Satcom and terrestrial specifications necessitates improvement to establish seamless connectivity.

### AI/ML in IMT 2030

The integration of Artificial Intelligence and Machine Learning into the realm of wireless communications is an area of vibrant research and development. In this context, the following key domains have garnered significant interest:

Advancing ML in the Physical Layer: This involves leveraging machine learning techniques to enhance various aspects of the physical layer, such as accurate channel estimation, precise symbol detection, error correction coding/decoding, and compressing Multiple-Input Multiple-Output (MIMO) Channel State Information (CSI).

Exploring ML in MAC and Higher Layers: Machine learning has also been extensively investigated for its potential in the Medium Access Control (MAC) and higher layers of wireless networks. Applications include multi-user scheduling, beamforming optimizations, traffic load balancing, and intelligent power control at cell sites to optimize coverage, among others.

Wireless networks generate copious amounts of data, thereby presenting opportunities for advanced data analytics empowered by ML technology. The seamless integration of these two domains holds the promise of automating network management, facilitating rapid fault detection and recovery, and reducing the need for manual intervention. Consequently, this transformative synergy is poised to drive revenue growth within the wireless communication industry.

### Integrated Sensing and Communication

Sensing and communication integration enables precise positioning with centimeter-level accuracy through wireless methods. This capability proves beneficial in various domains, particularly in industrial environments. In such scenarios, deploying multiple distributed transmitters within a facility allows for both highly accurate object localization and continuous monitoring. To support this application, a substantial bandwidth is necessary, which can be achieved through operating at extremely high frequencies, such as around 200 GHz. Additionally, this integrated approach facilitates the detection of activities, including human activity, through the utilization of machine learning techniques and the observation of changes in channel state information (CSI) within the environment.

## About Authors

| Author Name | Role | Organization | Email |
|---|---|---|---|
| Spandan Bisoyi | Research Scholar | IIT Hyderabad* | ee17resch11002@iith.ac.in |
| Murali Mohan Pasupuleti | Research Scholar | | ee21resch14005@iith.ac.in |
| Kiran Kuchi | Professor | | kkuchi@ee.iith.ac.in |
| Koteswara Rao | Senior Research Engineer | WiSig Networks | koti@wisig.com |
| Harish Kumar | | | harishkumar@wisig.com** |
| Pavan Reddy | Senior Research Engineer | | pavan@wisig.com |
| Smriti Kumari | Senior Research Engineer | | smriti@wisig.com |
| Sai Dhiraj Amuru | Principal Research Engineer | | asaidhiraj@wisig.com |

## About WiSig Networks

WiSig Networks Private Limited is a wireless innovator headquartered in Hyderabad India. WiSig was incubated in IIT Hyderabad Business Incubator in 2016, and it currently operates out of independent office in Financial District, Gachibowli, Hyderabad.

WiSig Networks builds state of art 5$^{th}$ generation wireless communication network technology, equipment, software and solutions. WiSig team consists of 100+ members including PhD graduates, research scholars, masters and graduate engineers from premier education and technology universities in India. WiSig Networks has so far filed over 160 patents, out of which 24 have been declared as Standard Essential Patents (SEPs).

WiSig Networks collaborates with semiconductor and processor companies to create products like 5G Radio Access Networks (RAN), Integrated Access Backhaul (IAB), licenses its designs and software to the interested Licensees including original equipment makers. WiSig also works with ecosystem partners like ODMs, EMS partners to manufacture and deploy wireless communication equipment, consumer and IoT devices.

## About IIT Hyderabad

Indian Institute of Technology Hyderabad (IITH) is a premier institute of science and technology established in 2008. IITH has been consistently ranked in the top 10 institutes in India for Engineering according to NIRF making it one of the most coveted schools for science and technology in the country.

IITH offers undergraduate programs in all the classical engineering disciplines, applied sciences, design, as well as several modern interdisciplinary areas. Students are given a flexibility to explore a broad set of areas, and potentially pursue a minor or double major in a discipline that is not their own. Students who wish to seek a deeper understanding of their own discipline are strongly encouraged to get involved in cutting-edge research with the help of a faculty to mentor them and earn an Honors in their own field.

The very foundation of IIT Hyderabad is based on research and innovation. The vibrant research culture is evident from the number of patents and publications that IITH produces consistently every year. IITH offers graduate programs at both a masters, and a doctoral level, in several diverse areas. There are separate programs for technology, design, science, and liberal arts. The MTech program is offered in various modes and durations to cater to the ever-growing need of postgraduate level professionals in the country.

IITH encourages and supports innovation and entrepreneurship at every stage. The minor program in entrepreneurship is open to all students of IITH, and many of the courses are offered by industrialists who are entrepreneurs themselves. There are various innovation centers and incubators that promote entrepreneurship and help mentor young innovators. IITH has been very successful in building tie-ups with leading academic institutions around the globe.

IITH enjoys a very special relationship with Japanese universities and industries that goes beyond academic and research collaborations. In fact, some of the iconic buildings in IITH campus will carry the signature of Japanese architecture. IITH creates a unique holistic ecosystem for education that offers interactive learning, a very flexible academic structure, cutting-edge research, strong industry collaboration, and entrepreneurship. This is an environment which enables students and faculty to translate their dreams into realities.

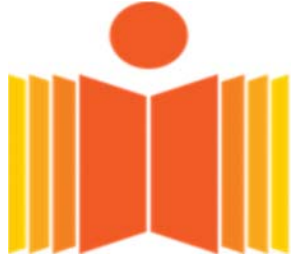

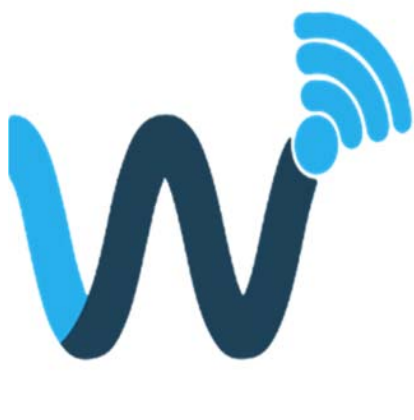

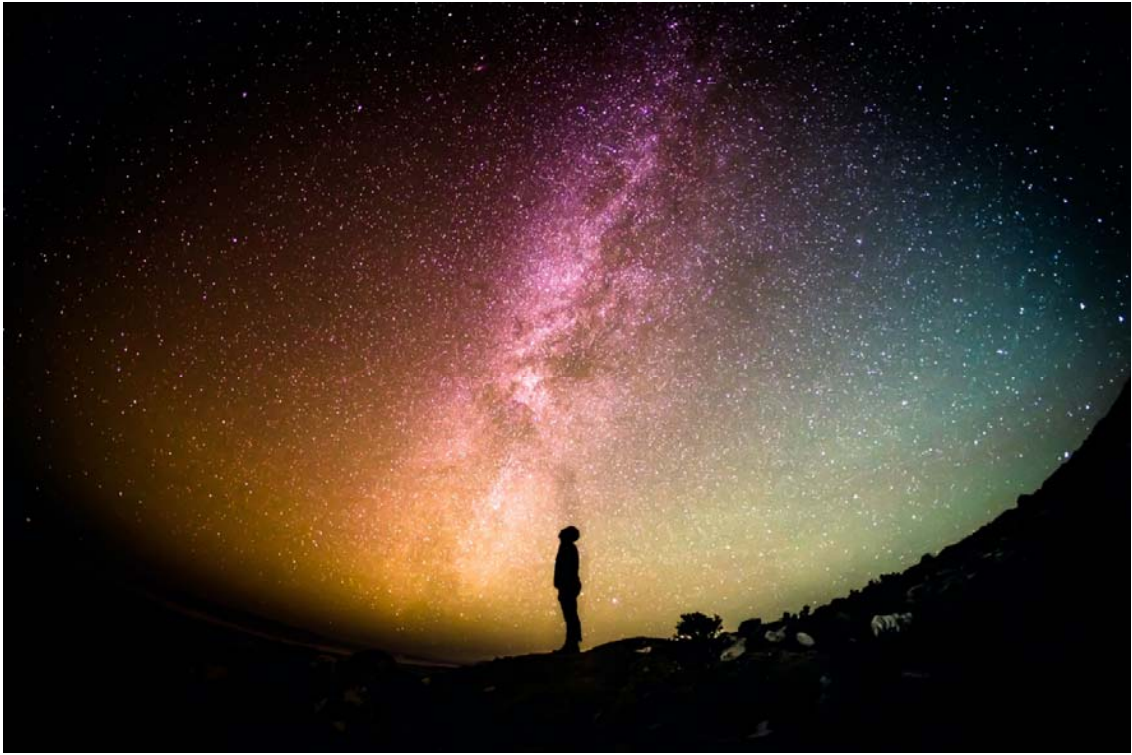